\documentclass[preprint2]{aastex631}

\usepackage{amsmath}
\usepackage{listings,jvlisting}
\usepackage[legacycolonsymbols]{mathtools}
\usepackage{ulem}
\usepackage{xcolor}

\usepackage{color}

\received{18 Jul 2023}
\accepted{03 Oct 2023}
\shorttitle{NICER X-ray Observations of Cir X-1}
\shortauthors{Tominaga et al.}
\graphicspath{{./}{figures/}}

\begin{document}

\title{X-ray spectral variations of Circinus X-1 observed with \textit{NICER} throughout an entire orbital cycle}

\correspondingauthor{Mayu Tominaga}
\email{tominaga@ac.jaxa.jp}

\author[0000-0002-1163-6077]{Mayu Tominaga}
\affiliation{Institute of Space and Astronautical Science, Japan Aerospace Exploration Agency,\\
3-1-1 Yoshinodai, Chuo-ku, Sagamihara, Kanagawa 252-5210, Japan}
\affiliation{Department of Astronomy, Graduate School of Science, The University of Tokyo,\\7-3-1 Hongo, Bunkyo-ku, Tokyo 113-0033, Japan}
\author[0000-0002-9184-5556]{Masahiro Tsujimoto}
\affiliation{Institute of Space and Astronautical Science, Japan Aerospace Exploration Agency,\\
3-1-1 Yoshinodai, Chuo-ku, Sagamihara, Kanagawa 252-5210, Japan}
\author[0000-0002-5352-7178]{Ken Ebisawa}
\affiliation{Institute of Space and Astronautical Science, Japan Aerospace Exploration Agency,\\
3-1-1 Yoshinodai, Chuo-ku, Sagamihara, Kanagawa 252-5210, Japan}
\affiliation{Department of Astronomy, Graduate School of Science, The University of Tokyo,\\7-3-1 Hongo, Bunkyo-ku, Tokyo 113-0033, Japan}
\author[0000-0003-1244-3100]{Teruaki Enoto}
\affiliation{Division of Physics and Astronomy, Graduate School of Science, Kyoto University, \\
Kitashirakawa, Sakyo-ku, Kyoto 606-8502, Japan}
\author[0000-0003-4799-1895]{Kimitake Hayasaki}
\affiliation{Department of Space Science and Astronomy, Chungbuk National University, Cheongju 361-763, Korea}
\affiliation{Harvard-Smithsonian Center for Astrophysics, 60 Garden Street, Cambridge, MA02138, USA}

\begin{abstract}
 Circinus X-1 (Cir X-1) is a neutron star binary with an elliptical orbit of
 16.6~days. The source is unique for its extreme youth, providing a key to understanding
 early binary evolution. However, its X-ray variability is too complex to reach a
 clear interpretation. We conducted the first high cadence (every 4 hours on
 average) observations covering one entire orbit using the \textit{NICER} X-ray
 telescope. The X-ray flux behavior can be divided into
 stable, dip, and flaring phases. The X-ray spectra in all phases can be described by a
 common model consisting of a partially covered disk black body emission and the line
 features from a highly-ionized photo-ionized plasma. The spectral change over the
 orbit is attributable to rapid changes of the partial covering medium in the line-of-sight and gradual changes of the disk black body emission. Emission lines of the H-like
 and He-like Mg, Si, S, and Fe are detected, most prominently in the dip phase. The Fe
 emission lines change to absorption in the course of the transition from the
 dip phase to the flaring phase. The estimated ionization degree indicates no
 significant changes, suggesting that the photo-ionized plasma is stable over the
 orbit. We propose a simple model in which the disk black body emission is partially blocked by
 such a local medium in the line-of-sight that has spatial structures depending on the 
 azimuth of the accretion disk. Emission lines upon the continuum emission are from the photo-ionized
 plasma located outside of the blocking material.
\end{abstract}


\section{Introduction} \label{s1}
Circinus X-1 (Cir X-1) is an X-ray binary system containing a neutron star
\citep{Margon1971} and a companion of an unidentified nature \citep{Jonker2007,
Johnston2001}. It is unique for its extreme youth among all the known X-ray binaries;
its association with the remnant of the supernova that produced the neutron star puts it
at an estimated age of $<$4600~yr \citep{Heinz2013}. It is also the first X-ray binary
from which the P Cygni profiles are clearly observed in the X-ray spectra
\citep{Brandt2000, Schulz2002}, indicating the presence of disk winds around the
neutron star. Such disk winds can contribute to the dissipation of the mass and angular
momentum of the system, thereby influencing the evolution of the binary system
\citep{Begelman1983}.

Despite the importance of Cir X-1 in many astrophysical contexts, the understanding of
the system has been hampered by the notoriously wild behavior of the X-ray flux across
various time scales. It has an orbital period of 16.6~days \citep{Kaluzienski+1976,
Jonker2007} and a significant eccentricity of $\sim$ 0.45
\citep{Jonker2007}. Across time scales longer than the orbital period, it shows drastic
flux changes over a few orders of magnitude from a milli-Crab to a super-Crab. In time scales shorter than the orbital period, the X-ray flux changes as a function of the
orbital phase in a regular \citep{Kaluzienski+1976} or an irregular \citep{Asai2014}
fashion. On the shortest time scales, it exhibits the twin kHz quasi-periodic
oscillations \citep{Tennant1987} and type I bursts  \citep{Tennant1986b,Linares2010}, both of which
are signatures of the accreting neutron stars with a low
magnetic field \citep{Klis2006}. Among these variations across different
time scales, the X-ray flux/spectral variations throughout the orbital phases are most
critical for understanding the accretion disk states around the compact object.

Many studies have been made on the X-ray spectral components and their changes over the
orbital period (e..g., \citealt{Shirey1996, Brandt1996, Iaria2001a, Iaria2002,
Schulz2002, Ding2006a, Ding2006b, DAI2007, Schulz2008, Ding2012, Schulz2020}). An
important work among them is the detection of the emission and absorption lines of
highly ionized metals in the high energy transmission grating (HETG) spectra with the
\textit{Chandra} X-ray Observatory, and a successful application of the photo-ionized
plasma model in a low flux state \citep{Schulz2008, Schulz2020}. Because
\textit{Chandra} observations are costly to cover the entire orbital cycle, the
high-resolution spectra were obtained only at limited orbital phases including the
periastron and apastron. An obvious next step is to cover the entire orbital phase with
a spectral resolution capable of detecting the photo-ionized plasma emission and
absorption features and constructing a model of the X-ray emitting objects. This is the
goal of this paper using the new data set obtained with the Neutron Star Interior
Composition Explorer (\textit{NICER}; \citealt{Gendreau2012, Gendreau2016}) telescope
covering an entire orbital period with a high cadence.

\medskip

We start with the observation and data reduction in \S~\ref{s2} and present the X-ray
light curve and spectral analysis in \S~\ref{s3}. We then discuss an interpretation of
the observed results with a simple model in \S~\ref{s4} and summarize the study in
\S~\ref{s5}.  Throughout the paper, we adopt the following; the distance to Cir X-1 of
9.4~kpc \citep{Heinz2015}, and the ephemerises of the periastron \citep{Nicolson2007}
given by
\begin{equation}
 \label{eq3-6}
  t(N)=\mathrm{MJD}~43076.32+(16.57794-0.0000401\times N)\times N,
\end{equation}
in which $N$ is the cycle number since MJD 43076.32.

\section{Observations and Data Reduction}\label{s2}
\textit{NICER} is an experiment located on the International Space Station (ISS) in 
near-Earth orbit at an altitude of $\sim$ 400~km. The X-ray Timing Instrument (XTI)
onboard \textit{NICER} consists of 56 sets of the X-Ray Concentrator optics (XRC;
\citealt{Okajima2016b}) and the Silicon Drift Detector (SDD;
\citealt{Prigozhin2012}). It has the capability of fast pointing with a robotic
mechanism. XTI's large effective area (600~cm$^{2}$ at 6~keV), moderate energy resolution
(an FWHM of 137~eV at 6~keV), fast absolute timing precision ($<$ 300~ns), and operational agility and flexibility opened an unexplored field for the timely X-ray
spectroscopy of variable sources, uncompromised by photon statistics and dynamic range. These features are particularly suited for X-ray binaries, in which fast variability of up to a few kHz and orbital flux changes are commonly
observed.

\textit{NICER} usually conducts observations consisting of a short ($\approx$500~s) snapshot with
many targets in one ISS orbit of 90 minutes. Cir X-1 was observed with 103 snapshots
totaling 61~ks (ObsID $=$\texttt{357801XX01}, where \texttt{XX}$=$02 to 17 and
3578011502, which we hereafter identify with a sequential number from 0 to 102) over an
orbital period with $\phi=$0.54--1.48 from 2020-08-01 to 2020-08-16
(MJD$=$59062--59078). The average revisit time is $\sim$ 3.9 hours, which corresponds
to $\sim$ 1 \% of the 16.6~day orbit.

We retrieved the level 2 products from the archive and processed them into the level 3
products using the \texttt{nicerl3} tools in \texttt{HEAsoft} version 6.31.1 and the
calibration database version 20221001. The level 3 products include the X-ray light
curves and spectra along with the redistribution matrix functions, the auxiliary
response files, and background spectra, which we used for the timing and spectroscopy
analysis. We modeled the background by SCORPEON
model\footnote{\url{https://heasarc.gsfc.nasa.gov/docs/nicer/analysis_threads/scorpeon-overview/}.}.
The SCORPEON model produces a background model fitted along with the source spectrum
including the following components; South Atlantic Anomaly (S), Cosmic Rays (COR), Polar and
Precipitating Electrons (PE), cOnstant terms (O), and Noise peak (N) of \textit{NICER}.

\section{Analysis}\label{s3}
We first explain the historical context of the \textit{NICER} observation and present the
X-ray light curve to define phases of the orbit based on flux changes (\S~\ref{s3-1}). We
then select several representative spectra in each phase to construct spectral models
and show that a common model can be applied to all phases despite apparent diversity (\S~\ref{s3-2}). Finally, we apply this common model to the entire data set
and present the change in the spectral parameters across its orbit (\S~\ref{s3-3})
separately for the continuum (\S~\ref{s3-3-1}) and line (\S~\ref{s3-3-2}) components.

\subsection{X-ray light curve}\label{s3-1}
\begin{figure*}[htb!]
\begin{center}
\centering 
 \includegraphics[keepaspectratio,width=2\columnwidth]{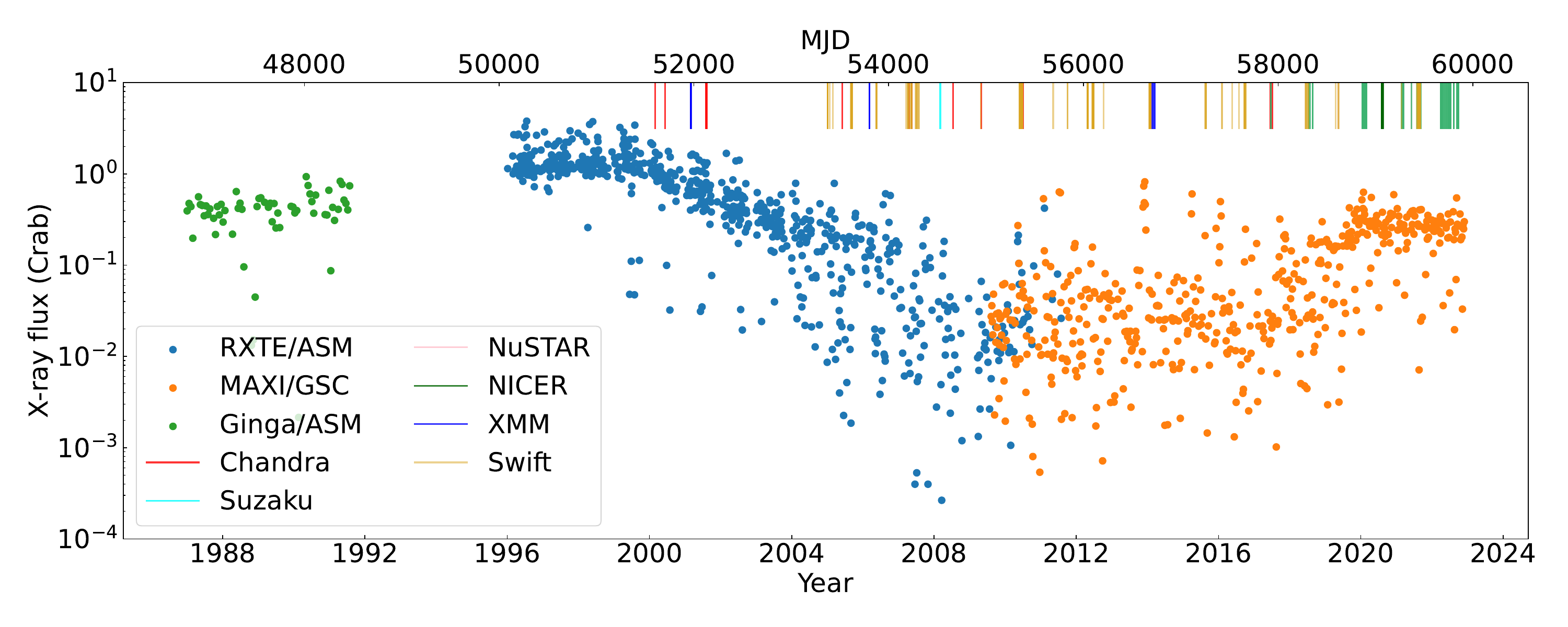}
 \caption{Long-term light curve of Cir X-1 observed with \textit{Ginga}, \textit{RXTE},
 and \textit{MAXI} all-sky monitors. Data points are sampled randomly for
 clarity. Vertical lines show the epochs of pointing observations with various X-ray
 telescopes.}
 \label{Fig7}
\end{center}
\end{figure*}

X-ray flux of Cir X-1 has been monitored for half a century. Light curves across three
decades from 1968 to 2001 are given in \citep{Parkinson2003}, which we extended for two
more decades up to 2023 in Figure~\ref{Fig7}. In the late 1990s and early 2000s, the
observed flux exceeded a Crab, during which P Cygni profiles of highly-ionized metals
were observed in X-ray grating spectra with \textit{Chandra}
\citep{Brandt2000,Schulz2002}. The flux declined to the 10 mCrab level throughout the
2010s, during which emission-dominated spectra were observed \citep{Schulz2020}. From
around 2018, the flux recovered to a few hundred mCrab at a similar level as in the
1980s. This is when repeated \textit{NICER} observations were made in its AO1--4 cycles
(green lines). We present the result of the data taken in the AO2 cycle in 2020.

\begin{figure}[htb!]
\centering 
 \includegraphics[keepaspectratio,width=\columnwidth]{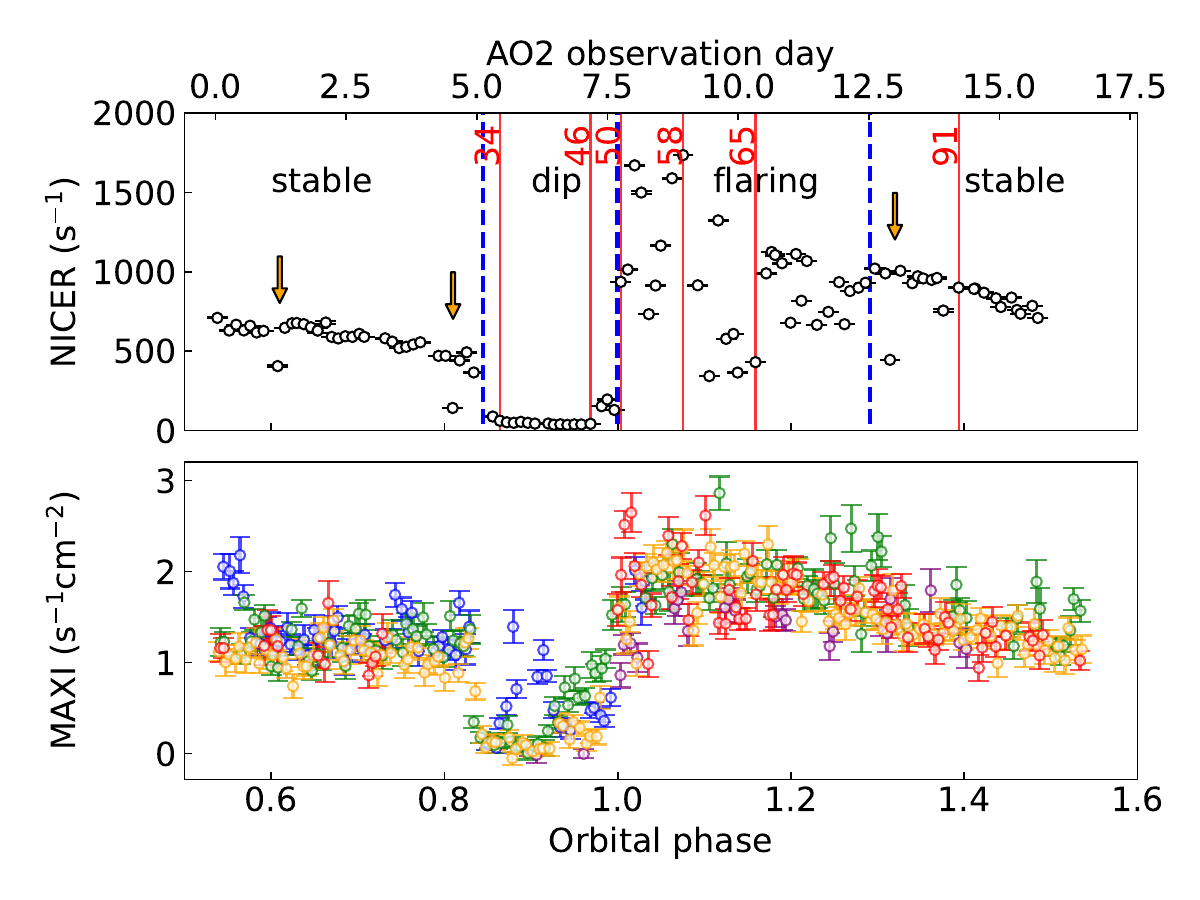}
 \caption{Light curves of the \textit{NICER} AO2 observation covering the orbital phase
 of 0.54--1.48 (top) and the \textit{MAXI} observation of the preceding five cycles of the
 same phase in different colors (bottom). In the top panel, the Bayesian-blocked phase
 boundaries are shown with blue dashed lines, while the data used for the spectral
 analysis representing each phase are shown with red solid lines with their snapshot
 numbers. Three orange arrows show the short dips.}
 \label{Fig1}
\end{figure}

After the flux recovered in 2018, the light curve folded by $P_{\mathrm{orb}}$
became more reproducible than that during the low-flux period, in which irregular flux
changes specific to an orbit were often observed \citep{Asai2014}. Figure~\ref{Fig1}
shows the \textit{NICER} light curve covering one orbit (top) compared to the
\textit{Monitor of All-sky X-ray Image} (\textit{MAXI}) light curve \citep{Matsuoka2009}
of five preceding orbits (bottom). Before the periastron at $\phi=1.0$, the flux drops
by an order of magnitude for $\Delta\phi \sim$ 0.15, which is often called ``long dips'' in the
literature (e.g., \citealt{Asai2014}). After the periastron, the flux suddenly increases
and becomes unstable for $\Delta\phi \sim$ 0.3. The flux decreases gradually thereafter. The \textit{NICER} data revealed these features at a higher significance and a
finer time resolution than any previous data. The sharp ingress and egress of the dips
at $\phi=$0.83 and 0.97, the rapid fluctuations at $\phi=$1.0--1.3, and small dips
at $\phi=$0.61, 0.81, and 1.32 (``short dips'') in a single orbit would not have been
detected without the \textit{NICER}'s flexible and photon-rich observations.

We divide the \textit{NICER} light curve into three phases using the Bayesian blocking
method \citep{Scargle2013}. The boundaries are shown with the dotted horizontal lines in
Figure~\ref{Fig1}.  We label them the ``dip'' ($\phi\sim$ 0.84--1.0), ``flaring''
($\phi\sim$ 1.0--1.29), and ``stable'' ($\phi\sim$ 0.54--0.84 and 1.29--1.48) phases. This
division is consistent with the \textit{ASCA} result \citep{Iaria2001a}, in which three
(dips, flaring, and stable) phases are recognized at $\phi=$ 0.78--0.97, 0.93--0.3, and
0.3--0.7 based on observations covering one orbit in March 1998.

\subsection{Modeling of X-ray spectra}\label{s3-2}
We constructed the X-ray spectra in the 0.9–-10.5~keV range for each of the 103
snapshots. This energy range was chosen to obtain a high signal-to-noise
ratio. As for the representative spectra, we selected six spectra in the three phases and
characterize them in \S~\ref{s3-2-1} (stable), \S~\ref{s3-2-2} (dip), and
\S~\ref{s3-2-3} (flaring). For the spectral modeling, we included the attenuation by photo-electric absorption \citep{Verner1996} from the interstellar medium (ISM) of the
solar abundance \citep{Wilms2000}. The spectra were rebinned to have at least 30 counts per energy bin and ensure the use of $\chi^2$ statistics.

\begin{figure*}[htb!]
 \centering 
 \includegraphics[keepaspectratio, width=2\columnwidth]{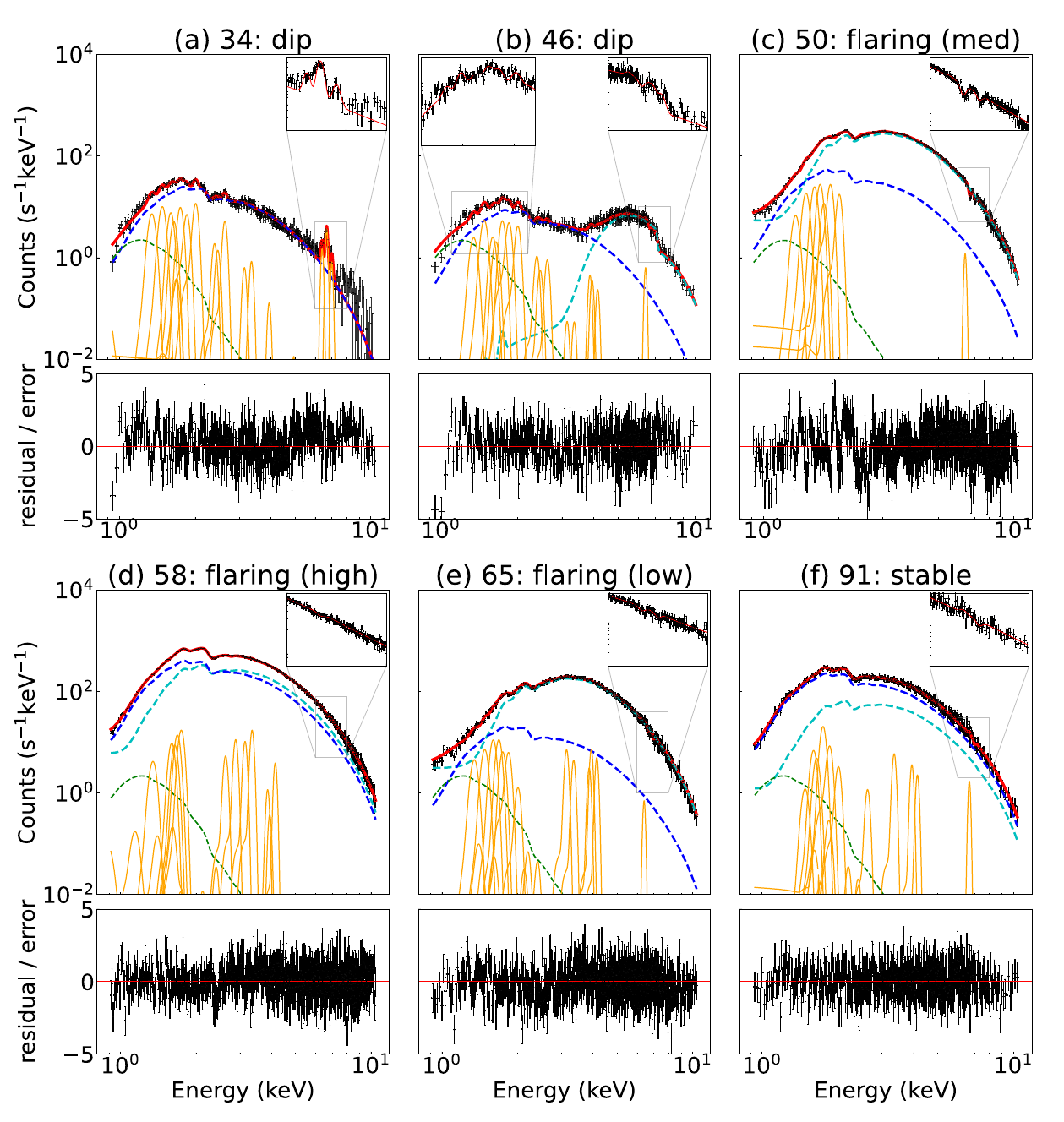}
 \caption{Spectra (black) and best-fit model in the top panels and the residuals to the
 fit in the bottom panels for the six snapshot spectra representing the dip (a and b), flaring (c,
 d, and e), and stable (f) phases. Different colors are used for different model
 components; red for total, cyan for covered disk black body, blue for uncovered disk
 black body, orange for Gaussian models, and green for soft excess. The insets give a
 close-up view of the energy bands with apparent spectral line features.}
 \label{Fig3}
\end{figure*}
 
\subsubsection{Stable phase}\label{s3-2-1}
We start with the stable phase. The continuum spectral shape shows only a gradual decrease in flux. This is shown in snapshot 91's data in Figure~\ref{Fig3} (f). The continuum emission can be modeled with a simple disk black body component
 \citep{Mitsuda1984,
Makishima1986} with two fitting parameters (the normalization and the innermost disk
temperature $T_{\mathrm{in}}$), partially covered by a neutral cloud
with two parameters (the covering fraction $f^{\mathrm{(cov)}}$ and the H-equivalent
column density $N_{\mathrm{H}}^{\mathrm{(cov)}}$) and interstellar absorption with a
fixed column density of 2.0$\times10^{22}$~cm$^{-2}$. \citet{Schulz2008} showed that
there can be several types of covering matter with different degrees of ionization along the line-of-sight. Our data did not allow us to distinguish these different types, thus we only assumed neutral
matter. We further added the Gaussian model to account for the Fe K fluorescence line at 6.4
keV.

\subsubsection{Dip phase}\label{s3-2-2}
We next investigate the dip phase. The spectral shape changed in the middle of the
dip phase. 
During the period of the first ten data points (hereafter ``the early dip'' period), the spectral shape exhibits one peak, which we represent with snapshot 34 (Fig.~\ref{Fig3}a). In contrast, during the period of the remaining six data points (hereafter "the late dip'' period), the shape exhibits two peaks, which we
represent with snapshot 46 (Fig.~\ref{Fig3}b). Despite the difference, both can be described with different parameters of the same model. The two peaks are, in
fact, reproduced by uncovered and covered components of the partially covered disk black
body emission. The spectra in this phase are also characterized by numerous emission
lines, most notably in highly ionized Fe, but also in lighter metals such as Mg, Si, and
S. We fit them with Gaussian lines likely originating from a photo-ionized plasma. The plasma emission should accompany continuum emission in the soft band. In fact, we did observe excess
emission in the softest end of the spectrum, which we phenomenologically represented with a power-law
model.

\subsubsection{Flaring phase}\label{s3-2-3}
There are significant flux variations in the flaring phase (Fig.~\ref{Fig1}). We chose three snapshots
representing high, medium, and low flux (snapshot 58, 50, and 65),
respectively). The flux varies by a factor of $>$5, but the spectral shape does vary as
much. We thus applied the same model (partially covered disk black body emission) and
we were able to model the spectra through variations in the covering fraction and column density of the partial covering component. Several Gaussian lines were added, both positive and negative in flux, to account for
the emission and absorption features, respectively.

\subsubsection{All the phases}\label{s3-2-4}
To summarize the result in \S~\ref{s3-2-1}--\S~\ref{s3-2-3}, despite the variety of the
spectral shapes, the spectra of the six selected snapshots were described well with the
same approach; the disk black body emission partially covered by neutral cloud and fully
covered by the ISM photo-electric absorption. Some minor differences remain, such as the
soft excess emission required for the late dip phase (Fig.~\ref{Fig3}b)
and the Gaussian lines observed either positive (emission) or negative (absorption)
in different phases.

We thus constructed a common model to encompass these minor differences. The soft excess
was required only in the dip phase, but we included this component in all phases with the
parameters fixed to the parameter values derived from the dip phase. This does not influence the
fitting in other phases, as the soft excess is overwhelmed by the uncovered disk black
body emission. We also added Gaussian lines at the energies of Ly$\alpha$ and He$\alpha$
transitions of H-like and He-like ions of Mg, Si, S, and Fe plus \ion{Fe}{1} K$\alpha$
fluorescence. Here, the He$\alpha$ line refers to all unresolved lines of (1s)(2p) or
(1s)(2s) $\rightarrow$ (1s)$^{2}$ transitions, while the Ly$\alpha$ line refers to (2p)
$\rightarrow$ (1s) transitions. We allowed both positive and
negative values in the fit to account for both emission and absorption. No significant Doppler
shift or broadening was observed in any of these lines, thus we fixed the line energy
shift and width to 0. The best-fit models shown in Figure~\ref{Fig3} are the results
of applying this common model.

\medskip

In the previous work of Cir X-1, the X-ray continuum spectra were often explained by a
combination of multiple spectral components of a black body, disk black body, and
Comptonized emission (e.g., \citealt{Shirey1999, Iaria2005, DAI2012}). We argue that a
single disk black body component is sufficient to explain the variety of 
spectra for the present data set. The partial coverage of the disk black body emission alone can make two
continuum peaks for the covered and uncovered parts. The addition of a black body component was not required in all spectra throughout the orbit. We note that the actual emission would be a mixture of emissions from the neutron
star surface, the boundary layer, and the accretion disk, but a
single disk black body component is sufficient to explain the data.

\subsection{Change of X-ray spectra}\label{s3-3}
Having constructed a common spectral model using the six selected snapshots, we apply
it uniformly to the spectra of all 103 snapshots. Although the fitting did not work for some individual spectra, the overall variation of the spectral parameters throughout the orbit could be characterized with this method. We first track the phase-dependent change of the
parameters of the continuum components (\S~\ref{s3-3-1}) and then those of the line
components (\S~\ref{s3-3-2}).

\subsubsection{Continuum}\label{s3-3-1}
\begin{figure}[htb!]
 \centering 
 \includegraphics[keepaspectratio,width=\columnwidth]{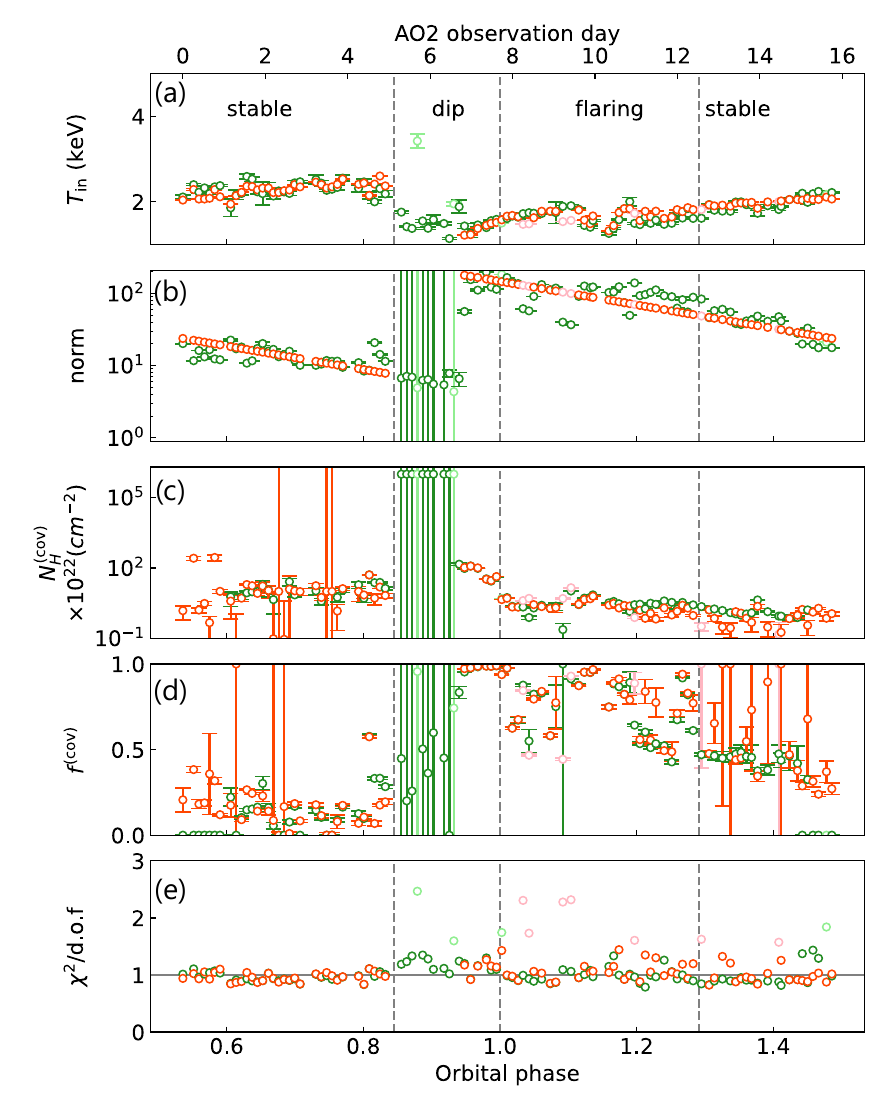}
 \caption{Phase-dependent change of the best-fit parameters for the continuum model: (a)
 $T_{\mathrm{in}}$ and (b) normalization of the disk black body emission, and (c)
 $N_{\mathrm{H}}^{\mathrm{(cov)}}$ and (d) $f^{\mathrm{(cov)}}$ of the partial covering
 material. (e) The reduced $\chi^{2}$ value of the fitting is also given. The green symbols represent the results without constraining the normalization parameter
values, whereas the red symbols, except for the early dip period, are those with the
normalization values constrained along the global trend.
Pale
 colors are for those with unacceptable fitting with a reduced $\chi^{2}$ greater than
 1.5, which we ignore when investigating overall spectral change.}
 \label{Fig4}
\end{figure}

The phase-dependent change of the spectral model parameters for the continuum components is given
in green symbols in Figure~\ref{Fig4}. The free parameters are the normalization and
$T_{\mathrm{in}}$ for the disk black body emission and $f^{\mathrm{(cov)}}$ and
$N_{\mathrm{H}}^{\mathrm{(cov)}}$ for the partial covering absorption.
To mitigate a degeneracy in the parameters space,
 firstly we fitted 
the data only above 4~keV, obtained the best-fit $T_{\mathrm{in}}$ values. We then fitted using all the available energy range with the fixed $T_{\mathrm{in}}$ values.

In the stable phase, the parameters change gradually. Some spectra do not require the
partial covering material, thus we fixed both $N_{\mathrm{H}}^{\mathrm{(cov)}}$ and
$f^{\mathrm{(cov)}}$ to null. In the flaring phase, we found that the fitting results
widely vary, in particular, for normalization. As this parameter represents the entire
disk luminosity, it is inconceivable that it changes as rapidly as the revisit time of
$\sim$4 hours.  We thus constrained its variation with a low-pass-filtered trend as
shown with red symbols in Figure~\ref{Fig4}. The rapid variation is now attributed
mostly to the changes in the partial coverage.

The trend of the normalization was further extrapolated backward to the late dip phase, which yielded acceptable fits. However, this did not work for the early dip phase. This is because of the different appearance of the spectral shape
as exemplified in the spectra from snapshot 34 and 46 (Fig.~\ref{Fig3}). In the late dip, we
observe both the partially covered and uncovered disk black body emission. In the early dip, we only observe the uncovered component, as the covered component is attenuated
too much to be visible in the \textit{NICER} energy band. The degeneracy between the following two cases
cannot be broken in the fitting: (a) a large value of the partial covering column or (b)
a small value of the disk normalization with a small covering column. The solution (b)
yields fitted results with a large scatter in the disk black body parameters
($T_{\mathrm{in}}$ and normalization), which is unlikely. We thus opted for (a) by
constraining the partial covering column to be $>10^{24}$~cm$^{-2}$ for the best-fit
parameter. For this reason We include large error estimates in the early dip phase in
Figure~\ref{Fig4}. The \textit{NICER} data do not allow us to track the trend of the
disk black body parameters during the early dip phase.

\subsubsection{Lines}\label{s3-3-2}
\begin{figure}[htb!]
 \centering 
 \includegraphics[keepaspectratio,width=\columnwidth]{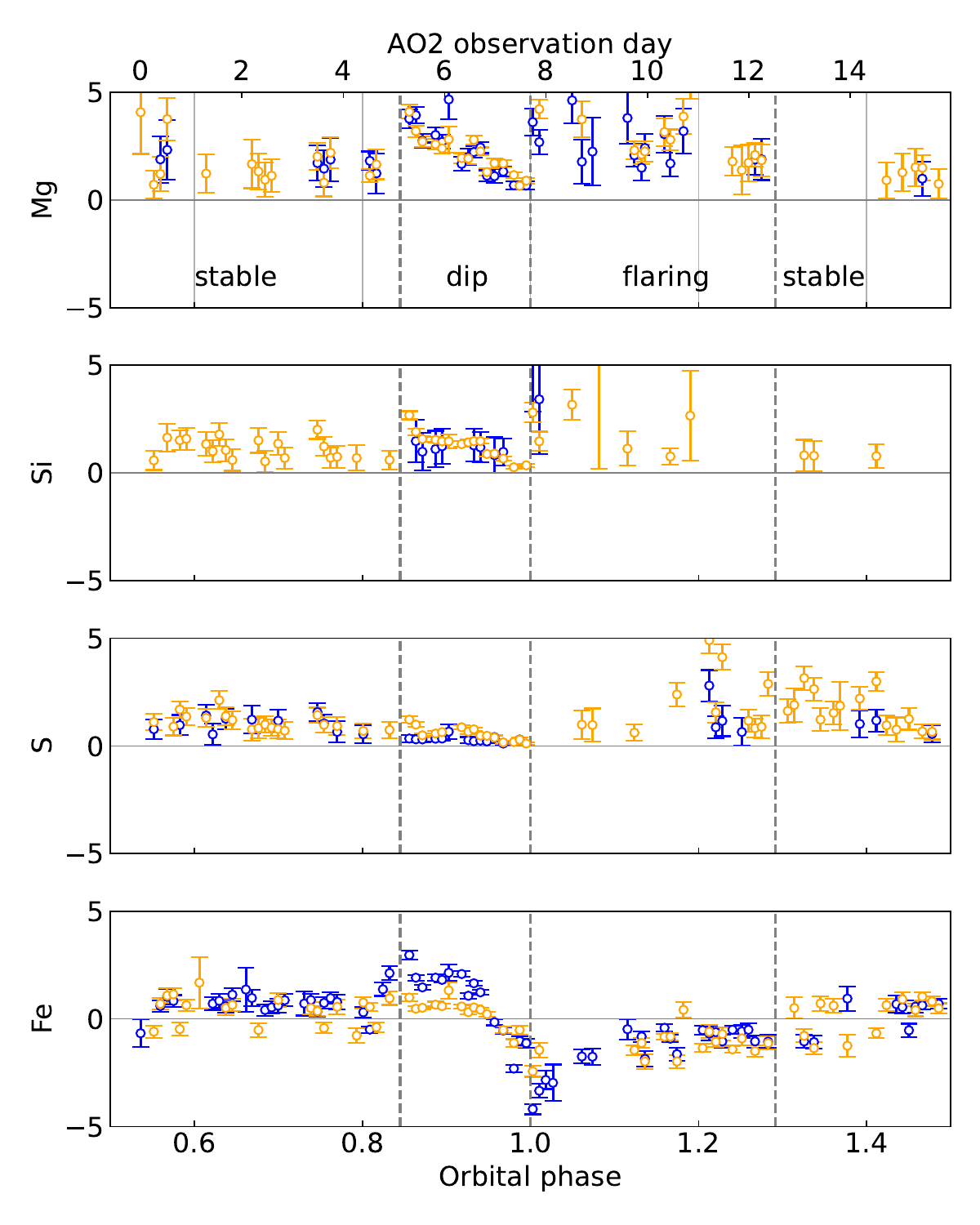}
 \caption{Normalization ($\times$10$^{-3}$~s$^{-1}$~cm$^{-2}$) of the line (positive for
 emission and negative for absorption) for He-like (blue) and H-like (orange) ions using
 the He$\alpha$ and Ly$\alpha$ lines of Mg, Si, S, and Fe. We omit the snapshots either with
 the reduced $\chi^{2}$ greater than 1.5 or the line normalization  consistent with
 zero.} 
 \label{Fig6}
\end{figure}

The phase-dependent change of the line normalization parameter is given in Figure~\ref{Fig6}
separately for Mg, Si, S, and Fe. Different colors are used for the H-like and He-like
ions.

For Mg, Si, and S, emission lines were detected in almost all the snapshots mainly
during the dip phase.  During the dip phase, their normalization decreases, particularly, in the late dip phase.

For Fe, a more stringent constraint was obtained, revealing changes between the
emission and absorption features. In the dip phase, the line appears as emission at the
beginning for the early dip, but the line normalization decreases in the late dip similar to the other elements. 
In the end, the line feature changed from emission to absorption, which
continues for the flaring phase and reverts back to emission in the stable phase.

\section{Discussion} \label{s4}
Based on the characterization of the entire data set in \S~\ref{s3}, we discuss the
interpretation. We first present the assumed geometry (\S~\ref{s4-1}), which we
rationalize by using a suite of observed results (\S~\ref{s4-2}), then we make some
quantitative assessment of the geometry (\S~\ref{s4-3}).

\subsection{Proposed Model}\label{s4-1}
Our assumed geometry is shown in Figure~\ref{Fig10}, in which the optically-thin plasma
extends away from the accretion disk. This picture is different from the conventional interpretation
of the ``(hot) accretion disk corona'', in which an optically-thin plasma exists on the
surface of the disk. The plasma is produced by being illuminated by the incident X-ray
emission from the surface of the neutron star and the accretion disk around it. The
plasma emits diffuse emission through radiative de-excitation and recombination. The
incident X-ray emission transmits through the plasma, in which absorption lines and
edges are imprinted by radiative excitation and ionization. The balance between them
depends on the ratio of the transmitted and diffuse emission determined by the geometry
in the line-of-sight and the incident X-ray flux.

We assume that a geometrically-flared optically-thick medium exists locally in the outer
part of the accretion disk. Such a structure is called the ``bulge'' in some literatures
\citep[e.g.,][]{White1982a}. We speculate that this is the accretion spot of the
material from the companion star, which is hinted at observationally
(e.g.,\citealt{Groot2001,Hynes2001,Hellier1996,Casares2003}) and envisioned
theoretically \citep{Armitage1998}. We do not delve into the detailed interpretation of what the covering media are. All we need is a localized structure that can provide a
complete or partial covering of the X-ray emission from the neutron star and accretion
disk, in which the covering column and fraction change as the sight line changes along
the binary orbit.

\subsection{Rationale}\label{s4-2}
We use four results to rationalize the proposed model. The first result is the presence of the disk
black body emission throughout the orbital cycle. Its parameters change only gradually on the orbital time scale. The flux variations are much faster than the orbital period primarily due to covering by
the medium outside the disk blackbody radiation region.
(Fig.~\ref{Fig4}).

The second result is the presence of the dip and flaring phases. The local covering medium should
have some density and clumpy structure depending on the disk's azimuth, so that it
creates changes in the line-of-sight as a function of the orbital phase. The X-ray flux
dip starts when the densest part of the covering medium comes in the line-of-sight.  The
total covering column decreases, resulting in the spectral changes between the early and the
late halves of the dip phase (\S~\ref{s3-2-2}). This dense material is broken into pieces in the tail part, which creates the rapid flux
variation during the flaring phase (\S~\ref{s3-2-3}).

The third result is the appearance of the Fe line feature either as emission or absorption in
addition to the disk emission (\S~\ref{s3-3-2}). The feature is made by highly-ionized Fe in the
photo-ionized plasma. When it appears as absorption, the incident disk emission
transmitted through the photo-ionized plasma surpasses the diffuse emission from the
plasma. When it appears as emission, the transmitted emission is insignificant by being
blocked in the line-of-sight, and the diffuse emission from the plasma is dominant.

The fourth result is the stability of the emission line strength compared to the total flux. The
total flux decreases by an order of magnitude in the dip phase (Fig.~\ref{Fig1}), while the emission
lines do not (Fig.~\ref{Fig6}). This indicates that most of the photo-ionized plasma is
not blocked by the local medium, unlike the disk black body emission.

\begin{figure}[htb!]
\centering
 \includegraphics[keepaspectratio,width=0.6\columnwidth]{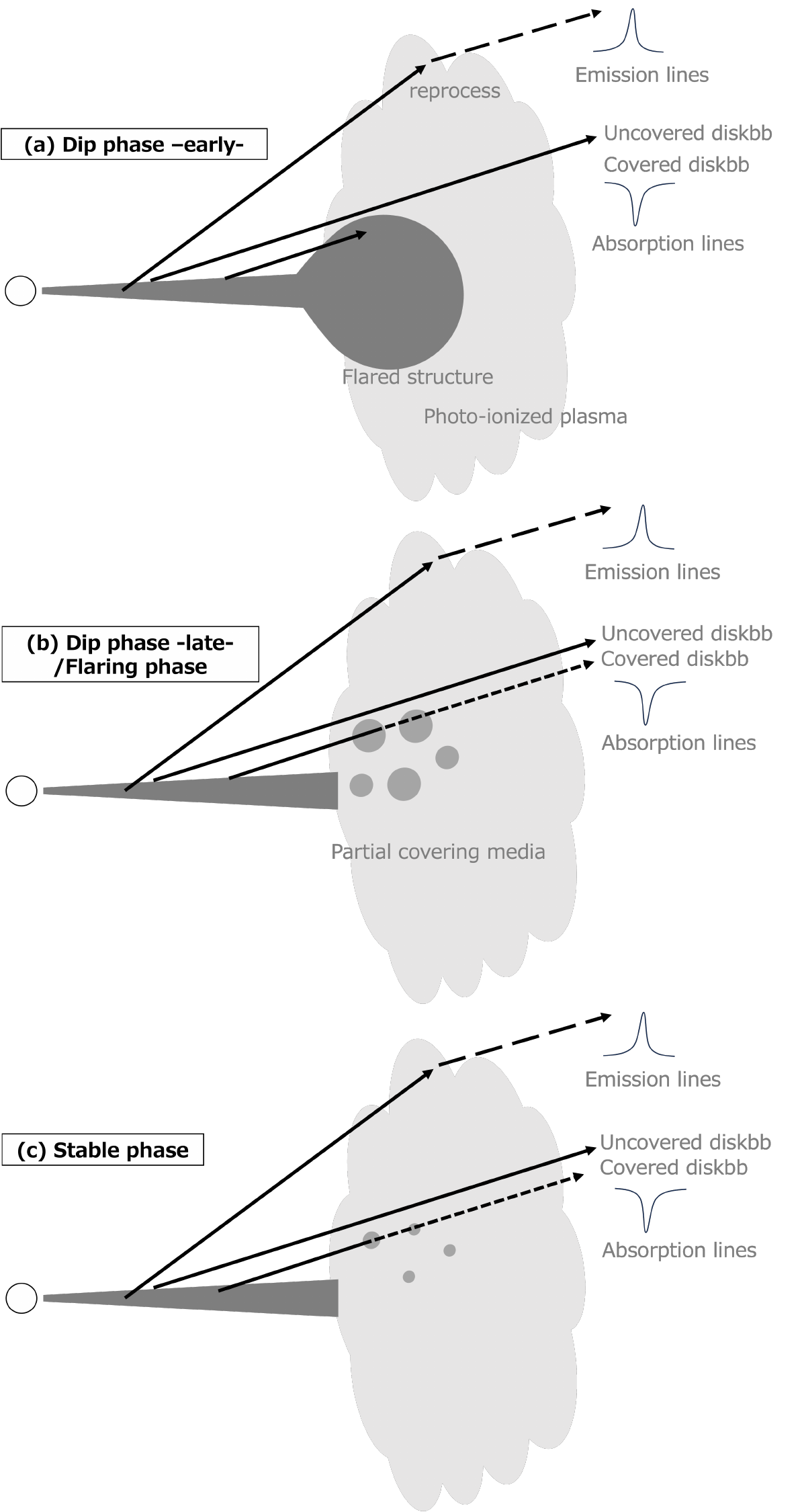}
 \caption{Proposed view of the system: edge-on views during the (a) dip (early), (b)
 dip (late)/flaring, and (c) stable phases, respectively. The local blocking medium
 has some spatial structures depending on the disk azimuth, making changes in the line
 of sight as a function of the orbital phase.}
 \label{Fig10}
\end{figure}

When all these are combined, we came to a model depicted in Figure~\ref{Fig10}. The
edge-on view for the early dip, late dip $+$ flaring, and stable phases
are shown in (a), (b), and (c), respectively.

\subsection{Assessments}\label{s4-3}
We make two quantitative assessments based on the geometry proposed in
Figure~\ref{Fig10}. One is the ionization parameter of the photo-ionized plasma
(\S~\ref{s4-3-1}) and the other is the system scale (\S~\ref{s4-3-2}).

\subsubsection{Ionization parameter}\label{s4-3-1}
We use the emission or absorption line features to derive ionization parameters. We calculate the intensity ratio between the Ly$\alpha$ and He$\alpha$
lines when both of them are detected. The emission line pair was detected for Mg in the
dip phase and for Fe throughout orbit, while the absorption line pair was detected
for Fe from the end of the dip phase to the flaring phase.

We employ the \texttt{xstar} code \citep{Kallman2001} to numerically estimate the
expected line ratio of the pairs. The code solves the radiative transfer equation for
the non-local thermal equilibrium condition for the charge and level populations in the
photo-ionized plasma by balancing radiative cooling and heating, ionization and
recombination, and excitation and de-excitation. It uses the two-stream approximation
solver in the inward and outward directions. The energy input (incident emission) is
given and the outputs (transmitted emission and diffuse emission inward and outward) are
calculated.

Because the \texttt{xstar} code is only for one-dimensional geometry, we approximated
the photo-ionized plasma as a plane-parallel slab. The slab is static, uniform, and
optically thin for the electron scattering with hydrogen density of $n_{\mathrm{H}}
= 3 \times 10^{14}$~cm$^{-3}$ and column density of $N_{\mathrm{H}} = 1
\times 10^{18}$~cm$^{-2}$. The incident emission has the shape of the disk black body
with representative parameters over the stable phase.

For the emission lines, they are mainly produced as a result of the recombination
cascades of e.g., Fe$^{26+}$ and Fe$^{25+}$ for Fe. The Ly$\alpha$ to He$\alpha$ ratio
is thus very sensitive to the ionization parameter $\xi=L/(n_{\mathrm{e}} r^{2})$,
in which $n_{\mathrm{e}}$ is the electron density and $r$ is the distance from the
incident source to the slab.  We took into account the effect of radiative excitation by
the incident emission, which is known to alter the line strength when a large fraction
of the incident emission is outside of the beam \citep{kinkhabwala2002} as in the dip
phase. The expected line ratio was calculated as a function of
$\log{\xi}$~(erg~s$^{-1}$~cm) over 0--5 in increments of 0.1 and a $\xi$ value was selected
that matches the observed ratio.

For the absorption lines, we followed the same procedure but evaluated the column density of Fe$^{25+}$
and Fe$^{24+}$ through the slab that is responsible for the absorption lines
respectively of Ly$\alpha$ and He$\alpha$ by absorbing the incident emission at each
transition energy. These column densities are related to the observed equivalent width
through the curve of growth. We selected the best $\log{\xi}$ value to reproduce the
observed ratio between the two lines.

\begin{figure}[htb!]
 \centering 
 \includegraphics[keepaspectratio,width=\columnwidth]{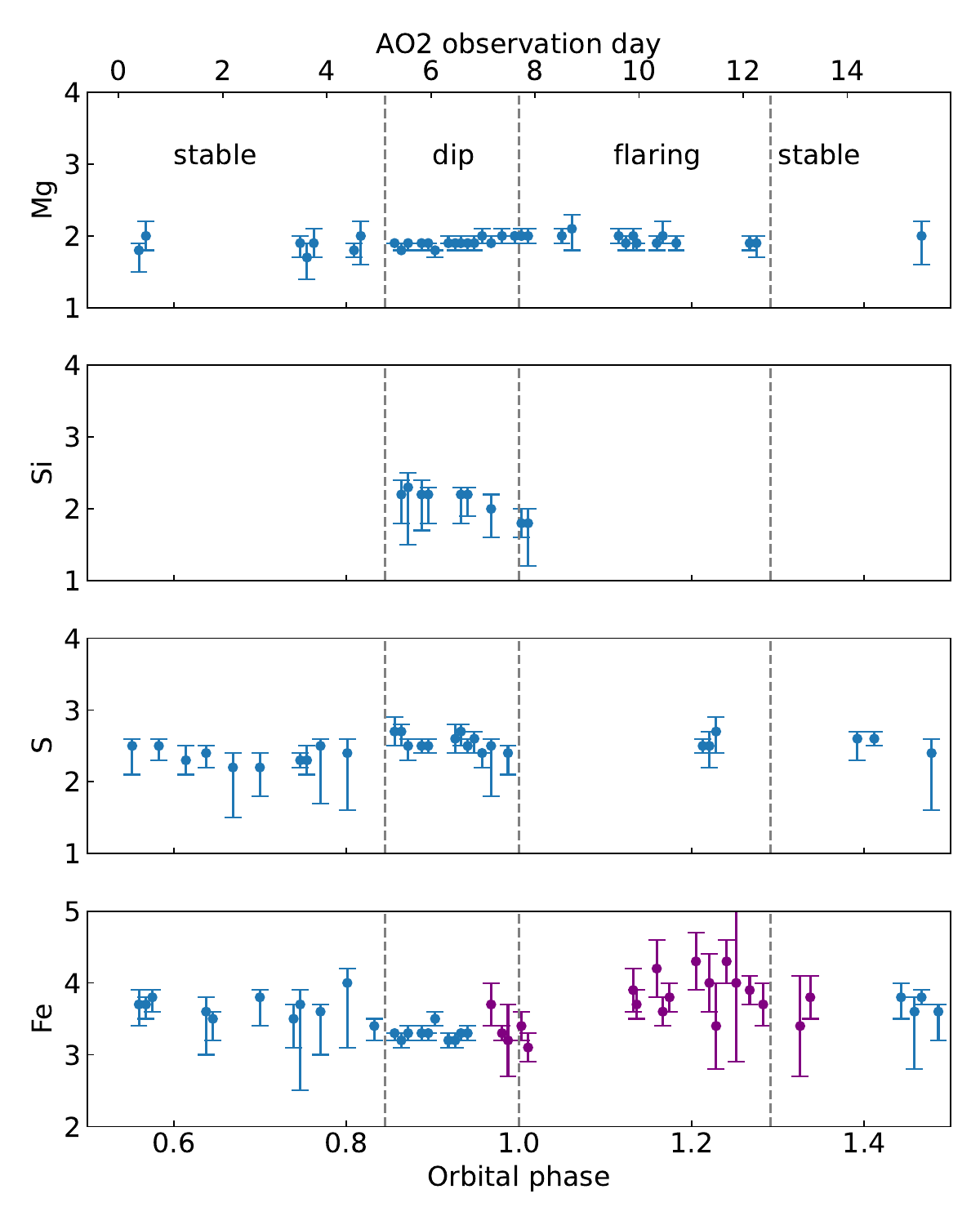}
 \caption{Best-fit ionization parameter ($\xi$) by comparing the observed and simulated
 line ratio of Ly$\alpha$ and He$\alpha$ lines for snapshots, in which both features are
 significantly detected. Blue symbols indicate values obtained using the emission line
 pairs and purple symbols indicate those using absorption line pairs.}
 \label{Fig9}
\end{figure}

The result is shown in Figure~\ref{Fig9}. The $\xi$ values using the Mg, Si, S emission,
Fe emission and Fe absorption pair are, respectively, $\sim 1.9\pm0.02$,
$\sim2.1\pm0.04$, $\sim2.5\pm0.02$, $\sim3.4\pm0.04$ and $\sim3.7\pm0.08$ for the means
and standard deviations. They differ to some extent, which indicates that the actual
photo-ionized plasma has a stratified ionization structure of varying $\xi$ as suggested
in \citet{Schulz2008}. However, they are stable across different phases. This suggests
that the photo-ionized plasma remains mostly unchanged throughout the orbital cycle.
The $\xi$ values using the Fe emission and absorption features match well, indicating
that the differences are created by changes in the viewing geometry and/or the
incident X-ray flux, not by changes in the photoionized plasma itself.

In the dip phase, it should be noted that the Fe line pairs changed from emission to
absorption in the middle at $\phi=0.95$ before this phase ends (Fig.~\ref{Fig6}). Since the emission from the photo-ionized plasma is considered stable over orbit, this
change should be created by an increase in the fraction of the incident emission through the
photo-ionized plasma, which contributes to the absorption lines. This is presumably made
by the increase of both the normalization of the disk black body emission and the
decrease of the partial covering fraction in the line-of-sight as shown in
Figure~\ref{Fig4}(d) and depicted in Figure~\ref{Fig10}.

\subsubsection{System scale}\label{s4-3-2}
The size of the system can be estimated by assuming the electron density
$n_{\mathrm{e}}$. The \textit{NICER} data are incapable of constraining this value, so we
used the estimate of the upper limit ($10^{15}$~cm$^{-3}$) given by the \textit{Chandra}
grating spectroscopy of the triplet lines of He-like ions \citep{Schulz2020}.

We estimate the size of the photo-ionized plasma. The inner radius $r_{\mathrm{in}} =
\sqrt{L_{\mathrm{X}}/(\xi n_{\mathrm{e}})}\sim10^{10}$~cm for the best-fit $\log \xi$
parameter for Fe emission and the observed disk luminosity of $L_{\mathrm{X}}$. The
$r_{\mathrm{in}}$ value is a lower limit as we used an upper limit for $n_{\mathrm{e}}$.

The outer radius ($r_{\mathrm{out}}$) is constrained by the observed
the maximum total amount of Fe absorption column by
\begin{equation}
 N_{\mathrm{Fe}i+}\sim f_{\mathrm{Fe}i+} A_{\mathrm{Fe}} n_{\mathrm{H}} (r_{\rm{out}}-r_{\rm{in}}),
\end{equation}
in which $N_{\mathrm{Fe}i+}$ is the observed absorption column of the Fe$^{i+}$ ions ($i
\in \left\{24,25\right\}$), $f_{\mathrm{Fe}i+}$ is the charge fraction of the ion among
all Fe, $A_{\mathrm{Fe}}$ is the assumed solar Fe abundance relative to H, and
$n_{\mathrm{H}}$ is the H density, which we approximate as $n_{\mathrm{H}} =
n_{\mathrm{e}}$. The Fe column is derived from the observed equivalent width through the
curve of growth. The charge fraction and the curve of growth are derived from an
\texttt{XSTAR} simulation, in which the column density of the slab
$\log{N_{\rm{H}}}$~cm$^{-2}$ was changed 15--25 in increments of 1 for the ionization degree
of $\log{\xi}=3.5$~erg~s$^{-1}$~cm. We used snapshot 50 and 65
(Fig.~\ref{Fig3} c and e, respectively) where the Fe absorption features were strong. For
both H-like ($i=25$) and He-like ($i=24$) ions, the result is in the range of
$r_{\mathrm{out}}-r_{\mathrm{in}} = 10^{9-10}$~cm. The photo-ionized plasma has some
structures for different $\xi$ values for different elements (\S~\ref{s3-3-2}). The part
contributing to the Fe features is relatively thin with
$r_{\mathrm{out}}/r_{\mathrm{in}}=1.1-2$.

This still yields strong emission lines from the plasma. \texttt{XSTAR} calculates that
\ion{Fe}{25} He$\alpha$ will be 6--120 $\times10^{36}$~erg~s$^{-1}$ for
$N_{\rm{H}}=10^{24-25}$~cm$^{-2}$, which is much larger than the observed value of
$1.7\times10^{35}$~erg~s$^{-1}$. This is the case in the dip phase when there is no
significant absorption features imprinted in the disk black body emission that cancels
the emission. Therefore, we consider that the volume filling factor of the photo-ionized
plasma is small in comparison to the \texttt{xstar} assumption of a spherical medium.

\medskip

We further compare the inner and outer radius of the photo-ionized plasma
($r_{\mathrm{in}}$ and $r_{\mathrm{out}}$) with those of the accretion disk
($R_{\mathrm{in}}$ and $R_{\mathrm{out}}$). In our spectral analysis, the normalization
of the disk black body emission is related to $R_{\mathrm{in}}$ as
$(R_{\mathrm{in}}/0.94)^2\cos{\theta}$, in which $\theta$ is the viewing angle. For a
complete edge-on $\theta=0$ and the normalization of 100 (the largest in the trend in
Fig.~\ref{Fig4}b when a large fraction of the disk is considered to be exposed), we
obtain $R_{\mathrm{in}}\sim$10$^{6}$~cm. The $R_{\mathrm{in}}$ value is close to a
typical value of a neutron star radius since the accretion disk is thought to stretch inward almost close to the neutron star surface (e.g., \citealt{Frank2002}).

We cannot constrain $R_{\mathrm{out}}$ from the present data. At least, it
must be smaller than the effective Roche lobe radius of the binary of $\sim 10^{10}$~cm
\citep{Begelman1983}. If we assume that a typical value for low mass X-ray binaries of 
$R_{\mathrm{out}}\sim10^{9-10}$~cm \citep{Jimenez-Garate2002} is applicable to Cir X-1,  $r_{\mathrm{in}} \gtrsim
R_{\mathrm{out}}$, suggesting that the photo-ionized plasma is located outside of the
accretion disk.

\section{Summary}\label{s5}
We present the result of \textit{NICER} observations covering an entire orbital period
of Cir X-1. Thanks to its operational flexibility and unprecedent large
collecting area with moderate energy resolution, we obtained
well-exposed spectra along the orbital phase with sufficient cadence, energy resolution,
and uniform quality. Using this unique data set, we obtained the following results and
implications.

\begin{enumerate}
 \setlength{\itemsep}{-1mm}
 \item The flux change over an orbit is repeated reproducibly by comparing to the
       \textit{MAXI} light curves of preceding orbits. It is divided into three parts;  dip,
       flaring, and stable phases.
 \item We constructed spectral models for representative spectra in each
       phase. Despite the apparent changes, we argue that they can be described
       by a common model with different parameters. 
 \item The continuum emission is explained by a variable disk black body component 
       over the entire orbit, with a minor invariable soft-excess component expected
       from the photo-ionized plasma. 
 \item The variety of the X-ray flux changes and the continuum spectral shape is
       mostly attributable to rapid changes of the partial covering material in the
       line-of-sight and gradual changes of the accretion disk temperature and
       luminosity.
       \begin{enumerate}
	\item The dip phase, which occupies $\sim$15\% of the orbit before and at the
	      periastron, is characterized by a sudden decrease in the X-ray flux.  It
	      is attributable to a sudden increase in the absorption column of the
	      partial covering material in the line-of-sight provided that the disk
	      black body emission gradually changes.
	\item The following flaring phase for $\sim$30\% of the orbit is characterized by rapid changes of the X-ray flux, which is mostly attributable to rapid
	      changes of the covering fraction of the partially covering medium in the
	      line-of-sight.
	\item The stable phase is characterized by a gradual decrease of the X-ray
	      flux. This is governed by the gradual change of the disk black body emission.
       \end{enumerate}
 \item Upon the continuum emission, we found emission and absorption features from the
       H-like and He-like ions of Mg, Si, S, and Fe. They are presumably from 
       the extended photo-ionized plasma. They are most evident during the dip phase when the
       continuum emission of the disk black body is most attenuated. The Fe line feature
       turns from emission to absorption in the middle of the dip phase.
 \item Using a pair of Ly$\alpha$ and He$\alpha$ lines respectively of H-like and
       He-like Mg, Si, S, and Fe, we derived the change of the ionization parameter
       as a function of the orbital phase. The ionization parameter value remains stable, regardless of whether the
       pair appears in emission or absorption, suggesting that the photo-ionized plasma
       remains almost unchanged over an orbit.
 \item We proposed an interpretation to explain these findings. A photo-ionized plasma illuminated
       by the incident X-ray emission from the accretion disk and the neutron star is
       located radially outside of the accretion disk, not on its surface.
       The disk black body emission is blocked in
       the early dip phase due to a geometrically flared medium which is present only for a subtended small azimuth angle range of the accretion disk. The medium has some density and
       fragmentation structure, the variation of which generates the X-ray flux and spectral changes in
       the late dip and flaring phases.
 \item We estimated the scales of these structures. The photo-ionized plasma responsible
       for the Fe line features is estimated to be located adjacently outside of
       the accretion disk.
\end{enumerate}

The sampling frequency of the orbital spectral variation presented in this study is much denser than in the previous Cir X-1 studies. Yet, we presented a much simpler spectral model. We recognize
that a few spectra yielded bad fitting, but they do not alter the overall interpretation
presented here. A geometry was proposed to explain these findings. We argue that the
proposal will help disentangle the complex behaviors of this source observed in other
instruments at different flux ranges and energy ranges and guide toward a better
understanding of the unique binary system.

\begin{acknowledgements}
We appreciate careful proofreading by Frederick T. Matsuda at ISAS/JAXA and Juriko Ebisawa. 
M.\,T. is financially
supported by the JSPS grant number 21J20947. 
K.\,H. has been supported by the Basic Science Research Program through the National Research Foundation of Korea (NRF) funded by the Ministry of Education (2016R1A5A1013277 and 2020R1A2C1007219). 
This work was financially supported by the Research Year of Chungbuk National University in 2021.
This work made use of the JAXA's super-computing system JSS3.  
This research has made use of data and/or
software provided by the High Energy Astrophysics Science Archive Research Center
(HEASARC), which is a service of the Astrophysics Science Division at NASA/GSFC.  This
research also made use of data obtained from Data ARchives and Transmission System
(DARTS), provided by Center for Science-satellite Operation and Data Archive (C-SODA) at
ISAS/JAXA, including the MAXI data provided by RIKEN, JAXA and the MAXI team.
\end{acknowledgements}

\vspace{5mm}
\facility{NICER \citep{Gendreau2012}}
\software{
HEAsoft \citep{heasoft},
Xspec \citep{Arnaud1996},
XSTAR \citep{Kallman2004}
}

\bibliography{main}
\bibliographystyle{aasjournal}
\end{document}